\documentstyle[aps,prl]{revtex}

\newcounter{test}[section]

\newcommand{\li}{\limits}

\newcommand{\nn}{\nonumber\\}
\newcommand{\non}{\nonumber}

\newcommand{\Z}{Z\!\!\!Z}
\newcommand{\Ibb}[1]{ \mbox{\rm I\ifmmode\mkern
            -3.6mu\else\kern -.2em\fi#1}}
\newcommand{\ibb}[1]{\leavevmode\hbox{\kern.3em\vrule
     height 1.2ex depth -.3ex width .2pt\kern-.3em\mbox{\rm#1}}}
\newcommand{\C}{{\ibb C}}
\newcommand{\R}{{\Ibb R}}

\newcommand{\be}{\begin{equation}}
\newcommand{\ee}{\end{equation}}
\newcommand{\bea}{\begin{eqnarray}}
\newcommand{\eea}{\end{eqnarray}}

\begin{document}
\draft
\title{Composite Particles in the Theory of Quantum Hall
  Effect} 
\author{T. Asselmeyer and R. Keiper} 
\address{Institute of Physics, Humboldt University \\ Berlin, Germany}
\date{March, 13, 1998}
\maketitle
\begin{abstract}
  The formation of composite particles in the electron liquid under QHE
  conditions discussed by Jain in generalizing Laughlins many-particle
  state is considered by using a model for two-dimensional guiding
  center configurations. Describing the self-consistent field of
  electron repulsion by a negative parabolic potential on effective
  centers and an inter-center amount we show that with increasing
  magnetic field the ground state of so-called primary composite
  particles $\nu=\frac{1}{q}$, $q=1,3,5,\ldots $, is given for higher
  negative quantum numbers of the total angular momentum. By
  clustering of primary composite particles due to absorption or
  emission of flux quanta we explain phenomenologically 
  the quasi-particle structure
  behind the series of relevant filling factors $\nu=\frac{p}{q}$,
  $p=1,2,3,\ldots$.
  
  Our considerations show that the complicate interplay of
  electron-magnetic field and electron-electron interactions in QHE
  systems may be understood in terms of adding flux quanta $\Phi_0$ to
  charges $e$ and binding of charges by flux quanta.
\end{abstract}
\section{Introduction}
Since the experimental discovery of the integral and fractional
quantum Hall effect (IQHE, FQHE) \cite{KliDorPep:80,TsuStoGos:82} this
macroscopic quantum phenomenon is a real challenge in particular for
theoretical and mathematical physicists. Under QHE conditions
determined by interacting quasi-twodimensional charges of high
mobility at very low temperatures in presence of a strong
perpendicular magnetic field $B$ and a disorder potential $U$ electron
correlation has been discussed by using several approaches.

In 1983 Laughlin \cite{Lau:83} introduced an antisymmetric
many-particle wave function for the fermionic system characterising
the ground state of the correlated quantum liquid by the odd numbered
parameter $m=1,3,5,\ldots$ Laughlins trial wave function may be
written in the form
\begin{equation}
\chi_{\frac{1}{m}}=\prod\limits_{\stackrel{j,k}{(j<k)}} (z_j-z_k)^{m-1}\chi_1
  \label{lau1}
\end{equation}
with
\begin{equation}
\chi_1=\prod\limits_{\stackrel{j,k}{(j<k)}} (z_j-z_k)
\exp(-\frac{1}{4}\sum\limits_{i=1}^{N_e} |z_i|^2)  \label{lau2} \; . 
\end{equation}
Here $z_j=x_j+iy_j$ denotes the dimensionless electron positions in
units of the magnetic length $l_0=\sqrt{\frac{\hbar}{eB}}$, $N_e$ is
the total electron number. In this approach, interaction has been
taken into account for spin-polarized electrons in the lowest Landau
level $N=0$. The many-particle state (\ref{lau1}) contains $m$ fluxes per
electron. It is an exact solution for harmonic potentials
\cite{GirJac:83}, but 
an approximation for totally included Coulomb interaction. 

A generalization of Laughlins construction was given in 1989 by Jain
\cite{Jai:89} who proposed a FQHE state of the form
\begin{equation}
\chi_{\nu}=\prod\limits_{\stackrel{j,k}{(j<k)}} (z_j-z_k)^{m-1}\chi_p
  \label{jain1}
\end{equation}
where $\chi_p$, $p=1,2,3,\ldots $ denotes an unknown antisymmetric
function describing the $p$-th IQHE state. Then one composes the number of
fluxes per electron for the $p$-th Landau level $\frac{1}{p}$ with the
even number $(m-1)$ of fluxes in analogy to expression (\ref{lau1}).
By this way we obtain all relevant filling factors
\begin{equation}
\nu = \frac{p}{(m-1)p\pm 1} \; .
  \label{filling-jain}
\end{equation}
In (\ref{filling-jain}) the number $p$ can be interpreted as Landau
level index $N$ in the sense $p=N+1$, $N=0,1,2\ldots$, and electron 
number as well. The quantized filling factor (\ref{filling-jain})
$\nu=\frac{p}{q}=\frac{h{\cal N}_s}{eB}$ with the $2d$-electron
concentration ${\cal N}_s$, describes composite particles (CP) which
connect $p$ charges $e$ with $q$ flux quanta $\Phi_0=\frac{h}{e}$.

Generally, correlation in the considered many-particle system may be
investigated also by topological and field theoretical methods which 
leads to an
interpretation of $q$ and $p$ as winding numbers of the magnetic field
and the self-consistent gauge field, respectively \cite{AssKei:95}. Then,
interaction and structure formation are reflected by the linking
number $\nu=\frac{p}{q}$. 

The actual quasi-particle picture for the 
FQHE ground state makes use of attaching an even or odd number of fluxes
to a charge leading to composite fermions (CF) or bosons (CB), respectively.
The statistics transmutes between these composite particles.

In this paper we give a local approach to the QHE quasi-particle
picture using a simplified description by an effective potential on
the guiding centers $\vec{R}_i$ which are surrounded by appropriate 
one-electron states $(n,m)$.
Because of the famous works of Fock \cite{Foc:28} and Landau \cite{Lan:30},
it is well-known that the Schr\"odinger equation for a single electron
in a constant magnetic field leading to potentials proportional to
$r^2$ and $\frac{1}{r^2}$, has exact solutions. In dependence on
boundary conditions and calibration of the vector potential one finds
Landau functions or radial symmetric one-particle states. In
\cite{AssKei:95} we started with the latter to investigate the
dynamical electron-electron interaction including spin structures by
topological techniques. In this paper the relevant electron-electron
interaction is modelled by a mean-field potential and a cluster 
construction which aims at a detailed illustration of the quasi-particle
structure observed in QHE experiments \cite{DuStTsPfWe:93,WiRuWePf:93}. 

\section{Primary composite particles}
\label{sec:2}
As the starting point we consider spinless electrons decribed by Schr\"odingers
equation with the effective Hamiltonian
\begin{equation}
  H=H_{eB}+H_{eV}+H_{eU} \label{effect-Hamiltonian}
\end{equation}
where $B$, $V$ and $U$ denote the homogeneous magnetic field, a
selfconsistent electric potential (including partially ee-interaction) and
the disorder potential, respectively. Lateron we take into account the Fermi
or Bose statistics which result from electron correlation beyond the mean
field description. The eigenvalue equation for $H_{eB}$ in symmetric gauge
for the vector potential $A_\phi=\frac{1}{2}Br$ has the form
\begin{equation}
\left\{ -\frac{\hbar^2}{2M}\left[ \frac{1}{r}\frac{\partial}{\partial
    r}\left(r\frac{\partial}{\partial
    r}\right)+\frac{1}{r^2}\frac{\partial^2}{\partial\phi^2}\right] -
  \frac{i\hbar\omega_c}{2}\frac{\partial}{\partial\phi} +
  \frac{M\omega_c^2}{8}r^2\right\} \Psi(r,\phi) = E \Psi(r,\phi)
  \label{eigenv-eq}
\end{equation}
with the cyclotron frequency $\omega_c=\frac{eB}{M}$. Equation
(\ref{eigenv-eq}) is exactly solved by
\begin{equation}
\Psi_{nm}=\frac{1}{l_0}\tilde{R}_{nm}(r)\frac{1}{\sqrt{2\pi}}e^{im\phi}
\label{solution}
\end{equation}
and
\begin{equation}
  E_{nm}=\hbar\omega_c\left[n+\frac{|m|+1}{2}+\frac{m}{2} \right]
=\hbar\omega_c\left(N+\frac{1}{2}\right)
\label{eigenvalue}
\end{equation}
with the radial quantum number $n=0,1,2,\ldots$ and the magnetic quantum
number $m=-\gamma,\ldots,0,\ldots,N$; $\gamma$ denotes the degree of the
degeneracy of the Landau level $N$ for the finite quantum system. The
well-known radial functions $\tilde{R}_{nm}(r)$ containing polynomial and
exponential factors are related to the guiding center at $\vec{R}=0$. In the
following we consider the many-particle system in the parameter space of
guiding centers forming a $2d$-configuration $\{\vec{R}_i\}$. 

In a first step the one-particle Hamiltonian $H_{eV}$ has to be determined.
For an effective center in the electron layer of thickness $l_d$ the
repulsive ee-interaction between the states (\ref{solution}) should be
modeled by the potential $V(r)$ of an average charge $e$ on a homogeneous
cylinder with radius $l_r(B)$. Solving Poissons equation for the center at
$\vec{R}=0$
\begin{equation}
  \label{poisson}
  \frac{1}{r}\frac{d}{dr}\left(
    r\frac{d}{dr}\right)V(r)=-\frac{e}{\epsilon_0\pi l_r^2l_d}
\end{equation}
one obtains the solution
\begin{eqnarray}
  \label{poisson-solution}
  V(r)=-\frac{e}{4\pi\epsilon_0l_d}\left\{
    \begin{array}{ll}
      \left( \frac{r}{l_r} \right)^2 & \mbox{for}\qquad r<l_r \\
     ( 2\ln\left(\frac{r}{l_r}\right)+1) & \mbox{for}\qquad r>l_r
    \end{array}
\right.
\end{eqnarray}
which approximates magnetic field dependent ee-interaction in the
q2d-system decomposed in a negative quadratic part for $r<l_r(B)$ and
a logarithmic long distance part. Including the effective Hamiltonian
\begin{eqnarray}
  \label{effective-hamiltonian}
  H_{eV}=-\frac{M}{2}\omega_0^2r^2
\end{eqnarray}
with the parameter
\begin{eqnarray}
  \label{parameter-omega}
  \omega_0^2(B)=\frac{e^2}{2\pi\epsilon_0Ml_dl^2_r(B)}
\end{eqnarray}
representing the one-particle approximation in Schr\"odingers equation
(\ref{eigenv-eq}) we can note that the structure of wave functions
(\ref{solution}) remains unchanged whereas the degeneracy in
(\ref{eigenvalue}) is lifted according to
\begin{eqnarray}
  \label{eigenvalue2}
  \tilde{E}_{nm}=\hbar\tilde{\omega}\left[n+\frac{|m|+1}{2}+\frac{m}{2}
\frac{\omega_c}{\tilde{\omega}}\right]
\end{eqnarray}
with
\begin{eqnarray}
  \label{omega-tilde}
  \tilde{\omega}^2=\omega_c^2-4\omega_0^2=\frac{e^2B}{M^2}(B-B_c)\; .
\end{eqnarray}
Equation (\ref{omega-tilde}) holds for $B>B_c\sim 1 T$
where we used the rough estimation $l_r(B)\simeq 10^2\ldots 10^3\mbox{nm}$ 
and $l_d\simeq 10\mbox{nm}$. 
It determines a smaller renormalized energy scale
$\hbar\tilde{\omega}<\hbar \omega_c$ for the quasi-particles in
comparison to the bare particles. In case of vanishing ee-interaction
one has $\omega_0=0$, $\tilde{\omega}=\omega_c$ and the
energy e.g. of the degenerated, lowest Landau level $N=0$ is
$E_{00}=\frac{1}{2}\hbar \omega_c$ for $n=0$, $m=0$ and all negative values
of $m$.

Looking for the intriguing interplay between ee- and eB-interactions in the 
considered system we realize that the flux quantum number $|m|$ includes spin 
structures acording to $|m|=2s+1$ where $s=\frac{1}{2},\frac{3}{2}\ldots
\rightarrow |m|=2,4,\ldots$ and $s=0,1,2,\ldots\rightarrow |m|=1,3,5,\ldots$ 
reflect fermionic and bosonic quasi-particle states (\ref{solution}), respectively.

The splitted energy levels (\ref{eigenvalue2})
for $n=0$ and $m=0,-1,-2,-3,-4,-5,\ldots$ are given in table \ref{tab:1}.
Table \ref{tab:1} shows, that with increasing magnetic field the
ground state will be determined by higher $|m|$ and the relative
influence of ee-interaction on the quasi-particle energy
decreases. The splitted energy levels are restricted by the
condition
\begin{equation}
  \tilde{E}_{nm}> 0 \label{condition1}
\end{equation}
which means that only a definite number of levels $\tilde{E}_{0m}$ with 
negative $m$ takes place in the confining potential which is caused by the 
magnetic field. If we fix for example the magnetic field at $B=B^*$ and the 
parameter value $\omega_0(B^*)$ at
\begin{equation}
  \left(\frac{\omega_c}{\tilde{\omega}}\right)^* = \frac{5}{4}
\end{equation}
then it holds that all levels down to
\setcounter{test}{\theequation}
\addtocounter{test}{1}
\setcounter{equation}{0}
\renewcommand{\theequation}{\arabic{test}\alph{equation}}
\begin{eqnarray}
  \tilde{E}_{0,-3} = \hbar\tilde{\omega}^*\left(2 - \frac{15}{8}\right)
  = \frac{1}{8}\hbar\tilde{\omega}^* > 0 
\end{eqnarray}
are allowed but
\begin{eqnarray}
  \tilde{E}_{0,-4} = \hbar\tilde{\omega}^*\left(\frac{5}{2} - 
  \frac{5}{2}\right) = 0 
\end{eqnarray}
and
\begin{eqnarray}
  \tilde{E}_{0,-5} = \hbar\tilde{\omega}^*\left(3 - \frac{25}{8}\right)
  = - \frac{1}{8}\hbar\tilde{\omega}^* < 0 
\end{eqnarray}
\setcounter{equation}{\value{test}}
\renewcommand{\theequation}{\arabic{equation}}

\noindent
are forbidden. For a higher magnetic field $B^*$ with e.g.
$\left(\frac{\omega_c}{\tilde{\omega}}\right)^*=\frac{7}{6}$ the levels
$\tilde{E}_{0,-4}$ and $\tilde{E}_{0,-5}$ are allowed but $\tilde{E}_{0,-6}$
is forbidden. Thus there exists a magnetic field dependent ground state with
a maximum value $|m|_{max}$ that fulfills condition (\ref{condition1}) for
the ideal system without disorder.

In order to discuss the important role
of disorder we introduce the new parameter $\tau_{coll}$ being the time
between two collisions of the quasi-particle with the disorder
potential $U$. In the real system with randomly distributed impurities
besides (\ref{condition1}) the cyclotron condition
\begin{equation}
\tilde{\omega}\tau_{coll}>1
  \label{condition2}
\end{equation}
has to be fulfilled. For a certain degree of disorder $\tau_{coll}$ has a
definite value $\tau^*$ leading to
\begin{equation}
  \tilde{\omega}\tau^*=|m|_{coll}
\end{equation}
where $|m|_{coll}>1$ is generally an irrational number which must be
replaced by the next lower integer. Therefore one has 
to distinguish between the cases
\setcounter{test}{\theequation}
\addtocounter{test}{1}
\setcounter{equation}{0}
\renewcommand{\theequation}{\arabic{test}\alph{equation}}
\begin{eqnarray}
  |m|_{coll} &>& |m|_{max}
\end{eqnarray}
and
\begin{eqnarray}
  |m|_{coll} &<& |m|_{max} \; ,
\end{eqnarray}
\setcounter{equation}{\value{test}}
\renewcommand{\theequation}{\arabic{equation}}

\noindent
where $|m|_{max}$ would be allowed in the ideal system without disorder
where only condition (\ref{condition1}) must be fulfilled. 
Obviously the real quasi-particle ground state has to be characterized by the 
optimum value $|m|_{opt}$ that is the lower integer value  of the two
numbers $|m|_{max}$ and $|m|_{coll}$
\begin{equation}
  |m|_{opt}=\min (|m|_{max},|m|_{coll}) \; .
\end{equation}

Resulting from our consideration we identify the module of the relevant
negative quantum number $m$ with the number of fluxes $q$ belonging to one
charge for a composite particle $\nu=\frac{1}{q}$. In case of an odd number
of fluxes attached to the reference state $|0,0\rangle$ one has Laughlin
quasi-particles in the sense of composite bosons. The index $\nu$ is used
here for the quasi-particle ground state and the filling factor as well.
With $\nu=1,\frac{1}{3},\frac{1}{5},\ldots$ we denote so-called primary
composite particles.

\section{Clustered composite particles}
Here we remember that the one-particle Hamiltonian
(\ref{effect-Hamiltonian}) with the effective potential part
(\ref{effective-hamiltonian}) accounts for ee-interaction only approximately
on short distances. There remains an unknown rest of Coulomb interaction
betwen primary CP. This correlation part should be responsible for cluster
formation in the relevant quantum liquid leading to Jain quasi-particles.
These $N$-body interactions may be discussed in a phenomenological picture
employing the transfer of flux quanta from the magnetic field $B$ to the CP
or in opposite direction.

Starting from the bosonic primary CP $\nu=1$ which combines one charge 
and one flux 
\begin{eqnarray*}
  \nu = \frac{p}{q}=\frac{1}{1}=1 \label{scheme0}
\end{eqnarray*}
a cluster of $p=2,3,4,\ldots$ CPïs is generated by the scheme
\begin{eqnarray}
\label{scheme1}
  & &\frac{1}{1}\oplus\frac{1}{1}\oplus\frac{0}{1}\longrightarrow
  \frac{2}{3}\\ & &\frac{1}{1}\oplus\frac{1}{1}\oplus\frac{1}{1}
  \oplus\frac{0}{2}\longrightarrow\frac{3}{5}\non \\ & &\ldots\quad
  \longrightarrow \left(\frac{1}{2}\right)^+ \; .\non
\end{eqnarray}
Here the symbol $\oplus$ stands for clustering and
$\frac{0}{1},\frac{0}{2},\ldots$ denotes binding flux quanta changing
from the $B$ field to the cluster.

If we start from other primary CP it holds
\begin{eqnarray}
  & &\nu =\frac{1}{3} \label{scheme2}\\
  & &\frac{1}{3}\oplus\frac{1}{3}\ominus\frac{0}{1}\longrightarrow 
  \frac{2}{5}\nn & &\frac{1}{3}\oplus\frac{1}{3}\oplus\frac{1}{3}
  \ominus\frac{0}{2}\longrightarrow\frac{3}{7}\non\\ & &\ldots
  \longrightarrow \left(\frac{1}{2}\right)^-\non
\end{eqnarray}
where the symbol $\ominus$ indicates the change of flux quanta from
the CP into the acting field $B$. Consequently one finds
\begin{eqnarray}
\label{scheme3}
  & &\frac{1}{3}\oplus\frac{1}{3}\oplus\frac{0}{1}\longrightarrow 
  \frac{2}{7} \\& &\frac{1}{3}\oplus\frac{1}{3}\oplus\frac{1}{3}
  \oplus\frac{0}{2}\longrightarrow\frac{3}{11}\nn & &\ldots
  \longrightarrow \left(\frac{1}{4}\right)^+\non
\end{eqnarray}
and furthermore
\begin{eqnarray}
  & &\nu =\frac{1}{5} \label{scheme4}\\
  & &\frac{1}{5}\oplus\frac{1}{5}\ominus\frac{0}{1}\longrightarrow 
  \frac{2}{9}\non\\& &\frac{1}{5}\oplus\frac{1}{5}\oplus\frac{1}{5}
  \ominus\frac{0}{2}\longrightarrow\frac{3}{13}\non\\ & &\ldots
  \longrightarrow \left(\frac{1}{4}\right)^-\non
\end{eqnarray}
and
\begin{eqnarray}
\label{scheme5}  
  & &\frac{1}{5}\oplus\frac{1}{5}\oplus\frac{0}{1}\longrightarrow 
  \frac{2}{11}\non\\& &\frac{1}{5}\oplus\frac{1}{5}\oplus\frac{1}{5}
  \oplus\frac{0}{2}\longrightarrow\frac{3}{17}\\ & &\ldots
  \longrightarrow \left(\frac{1}{6}\right)^+\non
\end{eqnarray}
From the scheme (\ref{scheme1})-(\ref{scheme5}) and table \ref{tab:1}
one sees that in dependence on $B$ ground state energies
$\tilde{E}_\nu$ for $\nu=\frac{p}{q}$ are lowered by absorption of
flux quanta whereas emission of fluxes leads to higher
$\tilde{E}_\nu$. The infinite series of $\nu$ converge from both sides
to the limits
$\left(\frac{1}{2}\right),\left(\frac{1}{4}\right),\left(\frac{1}{6}\right),\ldots$
describing composite fermions.

In our picture the IQHE is naturally included. From
\begin{eqnarray*}
  \nu = \frac{p}{q}=\frac{1}{1}=1 
\end{eqnarray*}
one has
\begin{eqnarray}
\label{scheme6}
  & &\frac{1}{1}\oplus\frac{1}{1}\ominus\frac{0}{1}\longrightarrow
  \frac{2}{1} \\& &\frac{1}{1}\oplus\frac{1}{1}\oplus\frac{1}{1}
  \ominus\frac{0}{2}\longrightarrow\frac{3}{1}\non\\ & &\qquad\ldots\qquad
  \longrightarrow N \rightarrow\infty \non
\end{eqnarray}
which demonstrates the well-known one-particle behaviour of electrons
on fully occupied Landau levels.
Fractionally values in the region $1<\nu<2$ are generated by
\begin{eqnarray}
\label{scheme7}
  1+\left(\frac{1}{3},\frac{2}{5},\frac{3}{7},\ldots
  \right)&\longrightarrow& \left(\frac{3}{2}\right)^- \\ 
  1+\left(\frac{2}{3},\frac{3}{5},\frac{4}{7}, \ldots\right)
  &\longrightarrow& \left(\frac{3}{2}\right)^+ \non  
\end{eqnarray}
in correspondence to fully and partially filled Landau levels.

Finally we remark that the cluster construction
(\ref{scheme1})-(\ref{scheme7}) claims primary composite bosons
$\nu=\frac{1}{1},\frac{1}{3},\frac{1}{5},\ldots$ which permits a deeper
insight into the condensation to fermionic QHE states e.g. the state
$\nu=\frac{1}{2}$ of the half filled lowest Landau level.

\section{Qualitative discussion of the ee-interaction}
After the remarkable coincidence between the phenomenological model used in
the previous sections and the observed conductivity fractions
\cite{DuStTsPfWe:93,WiRuWePf:93}, we are trying to motivate the relevant
interaction term (\ref{effective-hamiltonian}) in the effective Hamiltonian
(\ref{effect-Hamiltonian}). From a general point of view we describe a
plane of interacting electrons with fluctuating geometry. The macroscopic
Hall conductivity as an expectation value of the corresponding microscopic
current-current correlator is a topological invariant \cite{AssKei:95}. Now
the idea is to encode the topological information of the system into its
geometrical structure in order to determine an effective Hamiltonian
containing the main part of interaction.

For that purpose we consider a system of $N$ electrons under QHE conditions
on a surface $\C$. The construction of the configuration space is given in
the following way. At first we keep in mind that every single-particle wave
function admits a support which means an area with non-zero value. The
configuration of one fermion is simply the whole surface $\C$. In case of
two particles we fix the support of the wave function for one particle. By
Paulis principle the second particle must be located on a different point
with respect to the first one.  So we obtain the configuration space
$\C\times\C\backslash D$ where $D$ is the support of the other wave
function. We denote this configuration space by $C_2$. The many-particle
case is similar and the $N$-fermion configuration space $C_N$ has been
constructed by the same procedure. 

Now we consider the topology of $C_N$ to
describe all interactions of the many-particle system. In advance we note
that the space $C_2=\C\times\C\backslash D$ is homotopic to
$\C\times\C\backslash\{w\}$ where $w$ denotes the center coordinate of $D$.
Obviously this must be true for $C_N$ in the same sense.
Let $(z_1,z_2,\ldots ,z_N)\in C_N$ with $z_i\in\C$ be the coordinate vector
of the electrons. The guiding centers are denoted by $w_i\in\C$ for the
$i$-th fermion. Because of the principle of identical particles we can
change the order of the coordinates $(z_1,z_2,\ldots ,z_N)$ without any
effect on physical quantities. So finally we obtain the $N$-particle
configuration space $B_N=C_N/S_N$ which is the above constructed space $C_N$
apart from an arbitrary, symmetric permutation $S_N$ of $N$-particles. Such
spaces are extensively studied in the theory of knots and links
\cite{Law:90}, in the theory of iterated loop spaces
\cite{Seg:73} and in the singularity theory \cite{OrlSolo:80,Arn:90}. 

The set of complex, square-integrable
functions over $B_N$ forms the Fock space $F_N=L^2(B_N)$ of our system.
We emphasize that according to \cite{Law:90} the topological description of
magneto-electrons in $F_N$ is generated by the holomorphic functions
\begin{eqnarray}
\label{generator-coh}
f_N(z_1,\ldots ,z_N,w_1,\ldots ,w_N)=  
\sum\li_{k=1}^N\prod\li_{j=1}^k \frac{1}{(z_j-w_j)}=\sum\li_{k=1}^N f_{k,N}\; .
\end{eqnarray}
Thus we translate the topological information of $B_N$ encoded in $f_N$ into
a suitable 2-dimensional manifold formed by the particles itself. As shown
in \cite{AssKei:95} the torus $T^2$ is convenient to calculate the
fractional filling factor. On the other hand the topological information to
get out an effective Hamiltonian is given by the combinatorical structure of
the triangulation. The simplices of such triangulation are equilateral
triangles formed by three particles leading to a hexagonal configuration on
the surface. Therefore we choose $N=3$ and obtain the generator
\begin{eqnarray}
\label{generator-coh-3}
f_3(z_1,z_2,z_3,w_1,w_2,w_3)&=&\frac{1}{z_1-w_1}+\frac{1}{(z_1-w_1)(z_2-w_2)}+\\
& &+\frac{1}{(z_1-w_1)(z_2-w_2)(z_3-w_3)}\nonumber\; .
\end{eqnarray}
After the projection $B_3\to T^2$ where all three coordinates $z_1,z_2,z_3$
will be projected on one coordinate $z\in T^2$ we get
\begin{eqnarray}
\label{interaction-pot}
f_3(z,w_1,w_2,w_3)&=&\frac{1}{z-w_1}+\frac{1}{(z-w_1)(z-w_2)}+\\
& &+\frac{1}{(z-w_1)(z-w_2)(z-w_3)}\nonumber\; .
\end{eqnarray}
Although the projection loses some information about the topology of $B_N$
encoded in (\ref{generator-coh-3}) the main part is conserved in
(\ref{interaction-pot}). For the interpretation of this generator we can
only give an intuitive explanation derived from similar problems in
conformal field theory (CFT). There is a rich literature about the relation
between CFT and the FQHE (for instance
\cite{FlVa:94,Var:95}) including the relation to the Chern-Simons theory
\cite{Wit:89}. In \cite{Arn:90} Arnold investigated the structure of
interactions with singularity theory giving the answer to the question which
functions generate the main properties of such systems. As an example we
consider the set of harmonic functions including the Coulomb potential.
According to the theorem of Hodge
\cite{Hod:41} this set is isomorphic to the cohomology groups of deRham
\cite{deR:54,deRKod:50}\footnote{In (\cite{BoTu:95} p. 8) the cohomology of
$\R^4\backslash\R$ is calculated to state that the generator of the second
cohomology group is given by the Coulomb potential.} the
generators of which are interpretable as the interaction
potentials. As an example of this principle we consider the case of a
charged point particle over $\C$ with charge density $\delta(z)$, $z\in\C$.
From the viewpoint of physics we have to consider the Poisson equation
\begin{equation}
\triangle V(z)=\delta(z)\qquad\mbox{ in }\quad\C
\end{equation}
or we restrict the potential $V(z)$ to the set of harmonic functions given
by the solutions of
\begin{equation}
\label{laplace}  
\triangle \tilde{V}(z)=0 \qquad\mbox{ in }\quad\C\backslash\{0\}\; .
\end{equation}
From the mathematical point of view all solutions of (\ref{laplace}) are
given by the Poincare dual of the generator of
$H_1(\C\backslash\{0\},\Z)=H_1(S^1,\Z)=\Z$. Thus if we consider the
particles as holes of the space then the Poincare dual of the first homology
group represents the (harmonic) potential.

In order to calculate the effective interaction potential for the concrete
case of a triangular configuration we consider the barycenter coordinate $z$
with $|z|=r/l_0$, $l_0=l_0(B)$ and $arg(z)=\phi$ of the triangle instead of the
coordinates $z_i$, $i=1,2,3,\ldots$. Thus after the reduction to one
triangle we obtain for the potential $V(z)=V(r,\phi)$ from
(\ref{interaction-pot}) the following expression
\begin{eqnarray}
 V(z)=\frac{e}{l_0}\left[\frac{z^2+z(1-w_2-w_3)+w_2w_3+1}{(z-w_1)
({z}-w_2)(z-w_3)}\right] \, .
\end{eqnarray}
But in the equilateral triangle the distance between the barycenter and the
points $(w_1,w_2,w_3)$ is always equal and stationary. So we find the leading
term
\begin{equation}
V(z)=\frac{e}{l_0}|z|^2\, \exp(i\triangle\phi)\; .
\end{equation}
The most important
difference between our approach and the standard Coulomb interaction consists
in the phase $\phi$. The relevant phase
difference $\triangle\phi=\pm 2\pi/3$ is given by the angles between lines connecting
the barycenter with the vertex points. 
From the fact that the cohomology groups of the torus are generated by real
functions we have to consider the real part of the interaction term which
leads to
\begin{eqnarray}
  H_{eV} =  \cos(2\pi/3)r^2\Omega^2 = -\frac{1}{2}\Omega^2{r^2} 
\label{eV-interaction}  
\end{eqnarray}
and justifies the expression (\ref{effective-hamiltonian}) where the 
parameter $\omega_0^2(B)$ may be related to $\Omega^2(B)$. According to our
calculation this three-body potential is harmonic and lifts degeneracy of
the Landau levels.

The total ee-interaction may be written in the Form
\begin{eqnarray}
\label{final-interaction}
H_{ee}=H_{eV}+ W
\end{eqnarray}
where the unknown rest $W$ describes many-body correlations beyond the
short-range potential term (\ref{eV-interaction}). In section 3 this
part of ee-interaction is phenomenologically treated by a cluster 
construction.

\section{Summary and conclusions}
In this paper we presented a new explanation of composite
particles in the QHE theory. The crucial ee-interaction has been
simulated by (i) a repulsive effective potential characterized by the
parameter $\omega_0(B)$ responsible for primary CP and (ii) cluster
formation of composite particles connected with absorption or emission
of flux quanta.  The role of collisions between electrons and randomly
distributed impurities described by the parameter $\tau^*$, consists
in limitation of the existence of composite particles
$\nu=\frac{p}{q}$ with higher denominators.

The experimentally relevant quasi-particles $\nu=\frac{p}{q}$, $q$ -
odd, are composite bosons transmuting to composite fermions with
filling factors $\nu=\frac{1}{2},\frac{1}{4},\frac{1}{6},\ldots$
\cite{DuStTsPfWe:93,WiRuWePf:93}.

If the magnetic field $B$ increases the number of flux quanta per
electron also increases whereas the influence of ee-interaction
decreases for primary CP. For bosonic CP ee-interaction is responsible
both for formation and decomposition of clusters of primary CP. In
consequence of the complicate interplay between eB- and
ee-interactions the effective magnetic field $B^*$ in the quantum
liquid is expected to show characteristic minima \cite{DuStTsPfWe:93}.

The numerator in the filling factor $\nu=\frac{p}{q}$ can be
interpreted both as number $N$ of quasi-particles bounded in clusters
or in the sense of $N=n+1$ as the index of renormalized by
$\hbar\omega_c\rightarrow \hbar\tilde{\omega}$ Landau levels included
in electron correlation.

A geometric foundation of the phenomenological picture for $N$-body
interactions under QHE conditions can be given by generation of holomorphic
functions of the type (\ref{generator-coh}). A topological interpretation of
the generally fractional filling factor as a linking number has been given
in \cite{AssKei:95}. In a recent paper \cite{FrNaTsWi:97} Fradkin et al.
demonstrated how ee-interaction under QHE conditions may be discussed by a
flux-attachment procedure within an effective field theory for Pfaffian
states.




\begin{table}[htbp]
  \begin{center}
    \leavevmode
    \begin{tabular}{|rl|rcl|}\hline
      \multicolumn{2}{|c|}{energy levels splitted from}  &
      \multicolumn{3}{c|}{variation range} \\[-0.4cm]
      \multicolumn{2}{|c|}{\qquad the lowest renormalized Landau level\qquad} &
      \multicolumn{3}{c|}{\qquad of the ee-interaction parameter
        $\displaystyle\frac{\omega_c}{\tilde{\omega}}$\qquad} \\[0.3cm]
      \hline
      $\tilde{E}_{00}$ & =
      $\hbar\tilde{\omega}\left(\displaystyle\frac{1}{2}-0\right)$ \qquad LLL
      & $1<$ & $\displaystyle\frac{\omega_c}{\tilde{\omega}}$ &
      $<2+\eta$\qquad ,\quad $0<\eta \ll 1$ \\[0.3cm] \hline 
      $\tilde{E}_{0,-1}$ & =
      $\hbar\tilde{\omega}\left(1 -
      \displaystyle\frac{1}{2}\displaystyle
      \frac{\omega_c}{\tilde{\omega}}\right)$    
      & $1<$ & $\displaystyle\frac{\omega_c}{\tilde{\omega}}$ & $<2$ 
      \\[0.3cm] 
      $\tilde{E}_{0,-2}$ & =
      $\hbar\tilde{\omega}\left(\frac{3}{2} -
      \displaystyle\frac{\omega_c}{\tilde{\omega}}\right)$    
      & $1<$ & $\displaystyle\frac{\omega_c}{\tilde{\omega}}$ & $<\frac{3}{2}$ 
      \\[0.3cm] 
      $\tilde{E}_{0,-3}$ & =
      $\hbar\tilde{\omega}\left(2 -
      \displaystyle\frac{3}{2}\displaystyle
      \frac{\omega_c}{\tilde{\omega}}\right)$ 
      & $1<$ & $\displaystyle\frac{\omega_c}{\tilde{\omega}}$ &
      $<\displaystyle\frac{4}{3}$ \\[0.3cm] 
      $\tilde{E}_{0,-4}$ & =
      $\hbar\tilde{\omega}\left(\frac{5}{2} -
      2\displaystyle\frac{\omega_c}{\tilde{\omega}}\right)$    
      & $1<$ & $\displaystyle\frac{\omega_c}{\tilde{\omega}}$ & $<\frac{5}{4}$ 
      \\[0.3cm] 
      $\tilde{E}_{0,-5}$ & =
      $\hbar\tilde{\omega}\left(3 -
      \displaystyle\frac{5}{2}\displaystyle
      \frac{\omega_c}{\tilde{\omega}}\right)$
      & $1<$ & $\displaystyle\frac{\omega_c}{\tilde{\omega}}$ &
      $<\displaystyle\frac{6}{5}$ \\[0.3cm] 
      . & & & . & \\ . & & & . & \\ . & & & . & \\ \hline
      $|m|$ & $\longrightarrow\infty$ & &
      $\displaystyle\frac{\omega_c}{\tilde{\omega}}$ &
      $\longrightarrow 1^+$\\[0.3cm] \hline 
    \end{tabular}
  \end{center}
  \caption{quasi-particle energy levels for $n=0$ and 
    $m=0,-1,-2,-3,-4,-5,\ldots$}
  \label{tab:1}
\end{table} 
\end{document}